\documentclass[iop,numberedappendix]{emulateapj}

\usepackage[utf8]{inputenc}
\usepackage{amsmath}
\usepackage{amssymb}
\usepackage{graphicx}
\usepackage{hyperref}
\usepackage{times}

\begin{document}
\title{Black hole mass, jet power and accretion in AGN}
\author{Yong-Yun Chen,\altaffilmark{1}
          Xiong Zhang,\altaffilmark{1,$\dag$}
         Dingrong Xiong,\altaffilmark{1,3}
         Xiaoling Yu\altaffilmark{1}}
\altaffiltext{1}{Department of Physics, Yunnan Normal University, Kunming 650500,
China}
\altaffiltext{2}{Department of Physics, Yunnan Normal University, Kunming 650500,
China\\$^{\dag}$e-mail:ynzx@yeah,net.}
\altaffiltext{3}{National Astronomical Observatories/Yunnan Observatories, Chinese Academy of Sciences, Kunming 650011,
China}

\begin{abstract}
We study the relation between accretion, black hole mass and jet power in AGN, by using a large group of blazars detected by the Fermi Large Area Telescope and radio galaxies. Our main results are as follows. (i) The jet power of FSRQs and FRII-HEG depends on the black hole mass, which suggests that the FSRQs and FRII-HEG are in Radiation-Pressure Dominated regime. The jet power of BL Lacs and FRI-LEG depends on the accretion, which suggests that the BL Lacs and FRI-LEG are in the Gas-Pressure Dominated regime. (ii) We find that most of FSRQs and BL Lacs have $\rm{P_{jet}>L_{BZ}^{max}}$, which suggests that the Blandford-Znajek mechanism is insufficient to explain the jet power of these objects. (iii) The FSRQs are roughly separated from BL Lacs by the Ledlow-Owen's dividing line in the $\rm{\log P_{jet}-\log M}$ plane, which supports the unified scheme of AGN. (iv) The FSRQs and BL Lacs have a clear division at $\rm{L_{bol}/L_{Edd}\sim0.01}$, and the distribution of Eddington ratios of BL Lacs and FSRQs exhibits a bimodal nature, which imply that the accretion mode of FSRQs may be different from that of BL Lacs. (v) We find a significant correlation between broad line luminosity and jet power, which supports a direct tight connection between jet power and accretion.
\end{abstract}

\section{INTRODUCTION}           %% first-level sections will be auto-capitalized
\label{sect:intro}
Blazars are the most extreme Active Galactic Nuclei (AGN) with a relativistic jet directed towards our line of sight (Urry \& Padovani 1995). The emission from the jets is highly boosted and dominates the AGN emission at all wavelengths due to a relativistic beaming effect (Sbarrato et al. 2014). Blazars are generally divided into two subcategories including BL Lac objects (BL Lacs) and Flat spectrum radio quasars (FSRQs). BL Lacs only have very weak or no emission lines, whereas FSRQs show strong emission lines. The classical division between FSRQs and BL Lacs is mainly based on the rest frame equivalent width (EW) of their broad emission lines. Particularly, if the EW is larger than $\rm{5\AA}$, the objects are classified as FSRQs, otherwise as BL Lacs (Urry \& Padovani 1995). The broad emission lines of FSRQs are produced by distant gas clouds in broad-line regions (BLRs) which are photoionized by the optical/UV continua radiated from the accretion disks surrounding massive black holes (Sbarrato et al. 2012). The difference of the broad line emission between FSRQs and BL Lacs may be attributed to their different central engines (Cavaliere \& D'Elia 2002; Cao 2002, 2003). Landt et al. (2004) introduced an analogous classification criterion and found that it is possible to discriminate between objects with intrinsically weak or strong narrow emission lines by studying the $\rm{[O_{II}]}$ and $\rm{[O_{III}]}$ EW plane. Abdo et al. (2010a) have classified blazars based on the synchrotron-peak frequency of the broadband SED. Ackermann et al. (2011) used the estimated value of $\rm{v_{peak}^{s}}$ to classify the sources as either a low-synchrotron-peaked blazar (LSP, for sources with $\rm{v_{peak}^{s}< 10^{14} Hz}$), an intermediate-synchrotron-peaked blazar (ISP, for $\rm{10^{14} Hz <v_{peak}^{s} < 10^{15} Hz}$), or a HSP blazar (if $\rm{v_{peak}^{s} >10^{15}}$ Hz).

Radio galaxies are often thought to be the parent population of blazars. The blazars are aligned to our line of sight, radio galaxies have their jets oriented at larger viewing angle (Urry \& Padovani 1995; Sbarrato et al. 2014). Radio galaxies are commonly divided into two subclasses including FRI and FRII based on their radio morphology (Fanaroff \& Riley 1974). FRI shows bright jets close to the nucleus, while FRII shows prominent hot spots are far from it. The FRI and FRII radio galaxies can be clearly divided in the host galaxy optical luminosity-radio luminosity ($\rm{M_{R}-L_{rad}}$) plane by a dividing line showing that radio power is proportional to the optical luminosity of the host galaxy (Ledlow \& Owen 1996). Wu and Cao (2008) found that the dividing line of the Ledlow-Owen relation for FRI/FRII can be roughly reproduced by using the maximum jet power available from ADAFs around kerr black hole as a function of black hole mass with a hybrid jet formation model (i.e., BP+BZ mechanism). However, what causes the FRI/FRII division is still unclear. Gopal-Krishna and Wiita (2000) suggested that the differences between FRI and FRII may be caused by the jets interact with the ambient medium with different physical properties. However, some authors thought that the differences between FRI and FRII may be the intrinsic difference of their central engines, such as different accretion models and the formation process of jets (e.g., Baum et al. 1995; Bicknell 1995; Reynolds et al. 1996; Meier 1999; Ghisellini \& Celotti 2001; Marchesini et al. 2004; Hardcastle et al. 2007). Ghisellini and Celotti (2001) found that the division between FRI and FRII actually reflected a systematic difference in accretion rate. The FRIs have generally lower accretion rate ($\rm{\dot{m}\leq0.01}$), while for FRII it was typically larger. Some authors also found that the FSRQs and BL Lacs can be divided in $\rm{\dot{m}\sim0.01}$ (see e.g. Narayan, Carcia \& McClintock 1997; Ghisellini, Maraschi \& Tavecchio 2009). They suggested that the FRI and FRII could correspond to different accretion models. Laing et al. (1994) suggested that FRIIs can be divided into high-excitation (HEG) and low-excitation galaxies (LEGs), which are also applied to some FRI. Buttiglione et al.(2009,2010) found that all HEG are FRII, while LEGs can be both FRI and FRII. Best \& Heckman (2012) suggested that HEG radio galaxies are fuelled at high rates through radiatively-efficient standard accretion disks, while LEG radio galaxies are fulled via radiatively inefficient flows at low accretion rates. Heckman \& Best (2014) suggested that most FRIs are LEG sources, and most FRIIs are HEG sources. Gendre et al. (2013) suggested that the LEG and HEG sources could be divided by the Ledlow and Owen's line. These results may suggested that the FSRQs vs FRII and BL Lac vs FRI unification can be easily reinterpreted physically as HEG/LEG because of the large population overlaps between HEGs and FRII and between LEGs and FRI.

In the frame of unification schemes, many authors have studied the unified scheme of FSRQs and FRII radio galaxies. Padovani and Urry (1992) found that the radio luminosity functions (RLFs) of FRII radio galaxies are consistent with the observed radio luminosity functions of FSRQs, which favor the unification of FRII radio galaxies and FSRQs. Analogously, the unified scheme of BL Lac objects and FRI radio galaxies has been extensively studied by many authors who used different approaches, such as comparison of spectral energy distributions (SEDs) in different wavebands and radio morphology (Owen et al. 1996; Capetti et al. 2000; Bai \& Lee 2001). Padovani and Urry (1991) derived that the RLFs of BL Lacs are consistent with the observed RLFs of FRI radio galaxies, which also favor the unification of BL Lacs and FRI radio galaxies. Since the launch of the Fermi satellite, we have entered in a new era of blazars research (Abdo et al. 2009, 2010). Up to now, the Fermi Large Area Telescope (LAT) has detected hundreds of blazars because it has about 20 folds better sensitivity than its predecessor EGRET in the 0.1-100 GeV energy rang. Sbarrato et al. (2014) found a good correlation between the broad line luminosity and radio luminosity, which suggested a direct tight connection between jet power and accretion rate. They got that the observational differences between blazar subclasses reflect differences in the accretion regime.

Blandford and Rees (1978) suggested that the radiation observed from blazars is dominated by the emission from relativistiv jets, which transport energy and momentum to large scales. However, the formation of jets is not still unclear in astrophysics (e.g. Meier et al. 2001). Many models have been proposed to explain the origin of jets. There are two main theoretical models of the formation of jet. One is that the rotational energy of the black hole is expected to be transferred to the jets by the magnetic fields threading the holes when assuming the black hole is spinning rapidly (i.e., the Blandford-Znajek (BZ) mechanism; Blandford \& Znajek 1977). Another one is that the jet can also be accelerated by the large-scale fields threading the rotating accretion disk (i.e., the Blandford-Payne (BP) mechanism; Blandford \& Payne 1982). In the BZ theory, Gosh \& Marek (1997) found that it is possible to find two regimes for standard disc, which may be dominated by the radiation pressure (RPD) or the gas pressure (GPD). Foschini (2011) suggested that the FSRQs and narrow-line Seyfert 1 are in RPD regime, while BL Lacs are basically in the GPD regime by using the data of 30 FSRQs and 9 BL Lacs from Ghisellini et al. (2010) and 4 $\rm{\gamma}$-ray narrow-line Seyfert 1 from Abdo et al. (2009a).

In this paper, we use a sample of blazars detected by Fermi Large Telescope (LAT) and radio galaxies  with measured jet kinetic power, black hole mass and Eddington ratio to explore the relationship between them. The samples are described in Section 2; the results and discussions are presented in Section 3; conclusions are in Section 4. The cosmological parameters $\rm{H_{0}=70kms^{-1}Mpc^{-1}}$, $\rm{\Omega_{m}=0.3}$, and $\rm{\Omega_{\Lambda}=0.7}$ have been adopted in this work.

\section{THE SAMPLE}
In order to study the relationship between FSRQs, BL Lacs, FRI and FRII radio galaxies, we tried to select the large group of clean Fermi blazars and radio galaxies with reliable redshift, black hole mass and jet kinetic power.
\subsection{The Fermi blazars}
We tried to get the large group of Fermi blazars with reliable broad line luminosity, black hole mass and jet kinetic power. We consider the sample of Xiong \& Zhang (2014). Xiong and Zhang (2014) collected a larger group of Fermi blazars with broad line luminosity, black hole mass and jet power. However, there are some blazars that we don't know the types in their samples. Hence, we only collected known types of blazars from their samples. Firstly, Xiong and Zhang (2014) considered the following samples to get broad line data: Cao \& Jiang (1999), Wang et al. (2004), Liu et al. (2006), Sbarrato et al. (2012), Chai et al. (2012), Shen et al. (2011), Shaw et al. (2012). Secondly, they considered the following samples to get black hole mass: Woo \& Urry (2002), Xie et al. (2004), Liu et al. (2006), Zhou \& Cao (2009), Zhang et al. (2012), Sbarrato et al. (2012), Chai et al. (2012), Leon-Tavares et al. (2011a), Shen et al. (2011), Shaw et al. (2012). Nemmen et al. (2012) estimated the jet power for a large sample of Fermi blazars. Xiong and Zhang (2014) collected the jet power from the Nemmen et al. (2012) for their sample. When the jet powers of blazars were not directly gotten from the Nemmen et al. (2012), Xiong and Zhang (2014) used the relationship between intrinsic $\gamma$-ray luminosity and jet power derived by Nemmen et al. (2012) to calculate the jet power. We use the broad line luminosity to get bolometric luminosity ($\rm{L_{bol}\approx10L_{BLR}}$) and use the ratio of bolometric luminosity and Eddington luminosity to calculate the Eddington ratio ($\rm{\dot{m}=L_{bol}/L_{Edd}}$, $\rm{L_{Edd}}=\rm{1.3\times10^{38}(M/M_{\odot})ergs^{-1}}$ for Fermi blazars. At last, we get a sample containing 159 clean Fermi blazars (112 FSRQs and 47 BL Lacs).
\subsubsection{The broad line luminosity}
Celotti et al. (1997) calculated the broad-line luminosity by scaling several strong emission lines to the quasar template spectrum of Francis et al. (1991), using Lya as a reference. Francis et al. (1991) set Lya flux contribution to 100, and the relative weights of the $\rm{H\alpha}$, $\rm{H\beta}$, MgII and CIV lines to 77, 22, 34, and 63, respectively. The total broad line flux is fixed at 555.76. The broad line luminosity value of each source has been derived by using these properties. When more than one line was presented, the approach that calculated simple average of broad-line luminosity estimated from each line was adopted. The rest of authors in our sample also used method proposed by Celotti et al. (1997) to calculate the broad line luminosity.
\subsubsection{The black hole mass}
Usually, there are three methods to calculate black hole mass, namely, the traditional virial black hole mass, the stellar velocity dispersion or the bulge luminosity, and the variation timescale. The black hole mass is estimated by traditional virial method for most of FSRQs in the sample (Woo \& Urry 2002; wang et al. 2004; Liu et al. 2006; Sbarrato et al. 2012; Chai et al. 2012; Shaw et al. 2012; Shen et al. 2011; Wu et al. 2011). For some sources, especially BL Lacs, the black hole mass can be estimated by the stellar velocity dispersion or the bulge luminosity (Woo \& Urry 2002; Zhou \& Cao 2009; Zhang et al. 2012; Sbarrato et al. 2012; Chai et al. 2012; Leon-Tavares et al. 2011a). For few sources, the black hole masses are estimated by the variation timescales (Xie et al. 1991, 2004, 2007). When more than one black mass is gotten for same objects in the sample, the average black hole masses are adopted.

We also notice the possible biases due to different methods were used for different samples. The black hole mass was calculated by different methods in our sample. Tremaine et al. (2002) suggested that the uncertainty in the $\rm{M-{\sigma}}$ relation is small, $\leq$0.21 dex. The uncertainty on the zero point of the line width-luminosity-mass relation is approximately 0.5 dex (Gebhardt et al. 2000; Ferrarese et al. 2001). McLure \& Dunlop (2001) suggested that the uncertainty of the $\rm{M_{BH}-M_{R}}$ relation is 0.6 dex. We assume that the average uncertainty of black hole mass is 0.44 dex. The estimated black hole masses from stellar dispersion velocity are $\rm{\log(M/M_{\odot})}$=8.29 (Mkn 421), 9.21 (Mkn 501), 9.03 (2005-489), 8.23 (BL Lac) (Woo \& Urry 2002). The traditional virial black hole masses are $\rm{\log(M/M_{\odot})}$=8.5 (Mkn 421), 9 (Mkn 501), 8.5 (2005-489), 8.7 (BL Lac) (Sbarrato et al. 2012). The estimated black hole mass from the variation timescales are $\rm{\log(M/M_{\odot})}$=7.6 (Mkn 421), 8.3 (Mkn 501), 8.1 (2005-489), 7.8 (BL Lac)(Xie et al.2004). We find that the largest difference between them is less than 1 order of magnitude.  Furthermore, we also find that the black hole masses were calculated by the variation timescales are a little lower than the black hole mass calculated by another two methods. However, these sources are very few, and have no significant impact on our results.

\subsubsection{The jet kinetic power}
The jet kinetic power is estimated by using the correlation between the extended radio emission and the jet power. According to search for X-ray cavities in different systems including giant elliptical galaxies and cD galaxies, Cavagnolo et al. (2010) obtained this tight correlation between the ``cavity'' power and the radio luminosity
\begin{equation}
\rm{P_{cav}}\approx\rm{5.8\times10^{43}(\frac{P_{radio}}{10^{40}ergs^{-1}})^{0.7}}\rm{ergs^{-1}}
\end{equation}
which is continuous over $\sim$6-8 decades in $\rm{P_{jet}}$ and $\rm{P_{radio}}$ with a scatter of $\approx$0.7 dex and $\rm{P_{jet}=P_{cav}}$. Meyer et al. (2011) used this formula to calculate the jet power for radio galaxies. Following Meyer et al. (2011), Nemmen et al. (2012) also adopted this formula to calculate the jet power for a large sample of Fermi blazars. They got the relationship between the intrinsic $\rm{\gamma}$-ray luminosity and the jet power
\begin{equation}
\rm{\log{P_{jet}}}=\rm{(0.98\pm0.02)\log{L_{\gamma}^{int}}+(1.6\pm0.9)}
\end{equation}
with a scatter of $\sim$0.64 dex. Nemmen et al. (2012) also obtained the relationship between observation $\rm{\gamma}$-ray luminosity and beaming factor, $\rm{f_{b}}\approx5\times10^{-4}(\rm{L_{obs}}/10^{49}\rm{ergs}^{-1})^{-0.39\pm0.15}$, where $\rm{f_{b}}$ is a beaming factor, $\rm{L_{obs}}$ is observation $\rm{\gamma}$-ray luminosity. The intrinsic $\rm{\gamma}$-ray luminosity L can be calculated by $\rm{L=f_{b}L_{obs}}$. When the jet powers of blazars were not directly gotten from the Nemmen et al. (2012), Xiong \& Zhang (2014) used the equation (2) to calculate the jet power. We should note the low significant of the $\rm{f_{b}}$ vs $\rm{L_{obs}}$ relation derived by Nemmen et al. (2012).

We also note that the jet power was calculated by two different methods, namely, the intrinsic radio luminosity (using a relationship with 0.7 dex scatter) and the intrinsic $\rm{\gamma}-ray$ luminosity (0.6 dex). We compare the jet power that calculated by equation (1) with that calculated by the equation (2) in Figs.1. Figure 2 shows that the jet power that calculated by equation (1) and that calculated by the equation (2) as a function of black hole mass. From figure 2, we can see that the distributions of jet power and black hole mass have no significant difference for our sample although the jet power was calculated by equation (2). From these figures, we can see that there is no bias in our results obtained within 3$\sigma$. Moreover, the bulk of our samples (about $60\%$) have jet powers that calculated by equation (1) and the minority of the sample (about $40\%$) have jet power that calculated by equation (2).
\begin{figure}
\includegraphics[width=8.6cm,height=8cm]{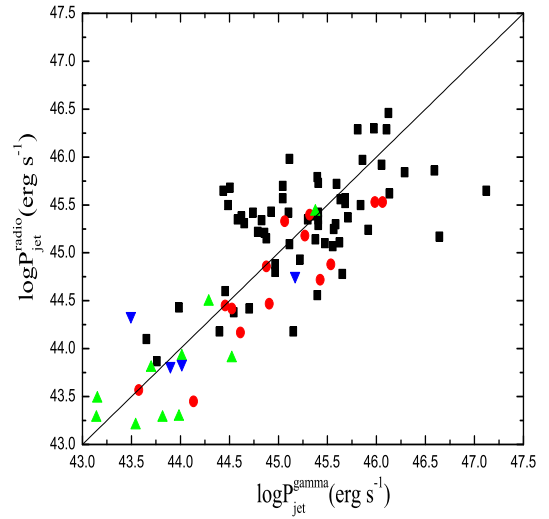}
\vspace{0pc}
\caption{The jet power calculated by equation (1) as a function of jet power calculated by equation (2) for FSRQs (black), LBL (red), HBL (green), IBL (blue). The black line is $\rm{\log{P_{jet}^{radio}}}$=$\rm{\log{P_{jet}^{gamma}}}$.}
\label{sample-fig2}
\end{figure}
\begin{figure}
\includegraphics[width=9cm,height=12cm]{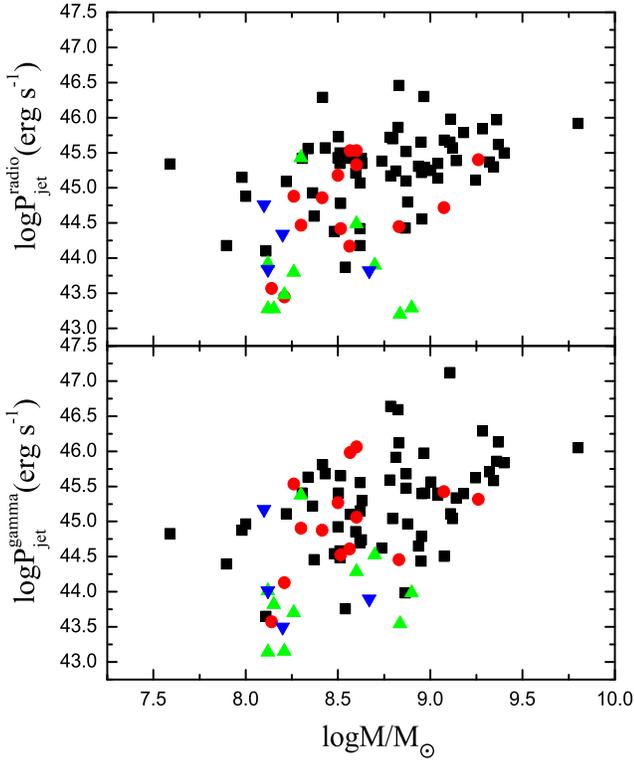}
\vspace{0pc}
\caption{The jet power calculated by equation (1) and jet power calculated by equation (2) as a function of black hole mass for FSRQs (black), LBL (red), HBL (green), IBL (blue).}
\label{sample-fig2}
\end{figure}
\subsection{The radio galaxies}
We tried to select the FRII-HEG and FRI-LEG radio galaxies with reliable redshift, black hole mass and jet power. Firstly, we consider the sample of Meyer et al. (2011). Meyer et al. (2011) estimated the jet power by the method of Cavagnolo et al. (2010) for a sample of radio galaxies. Buttiglione et al. (2010) have classified the sources as HEG, LEGs and broad line objects (BLOs) according to optical features. We cross-correlated Meyer's sample with Buttiglione's ample. We get the 14 FRI/LEG and 6 FRII/HEG radio galaxies. Secondly, we use the $\rm{M_{bh}-M_{R}}$ relation to calculate the black hole mass by using an average color correction of R$-$H=2.5 (Dunlop et al. 2003). At last, the narrow-line regions (NLRs) are believed to be photoionized by the radiation from the accretion disk and the narrow-line emission can be used to estimate the bolometric luminosity for radio galaxies. We calculate bolometric luminosity by using the relation proposed by Heckman \& Best (2014), namely, $\rm{L_{bol}=600L_{[OIII]}}$. We collect the narrow lines $\rm{[O_{III}]}$ from Buttiglione's ample and calculate Eddington ratio.
\section{RESULTS AND DISCUSSIONS}
In the BZ theory, the magnetic field of the disc is pushed toward the event horizon by the Maxwell pressure (Foschini 2011). For a standard disc, its magnetic field depends on the accretion rate and it is possible to find two regimes (Ghosh \& Abramowicz 1997). One refers to strong accretion disc, which is dominated by the radiation pressure (RPD). Ghosh and Abramowicz (1997) pointed out that the jet luminosity of the BZ mechanism in the radiation pressure dominated can be estimated by the following formula
\begin{equation}
\rm{L_{BZ,RPD}}=\rm{2\times10^{44}(\frac{M}{10^{8}M_{\odot}})j^{2}ergs^{-1}}
\end{equation}
where M is the black hole mass, and j is the dimensionless angular momentum. The other is the standard disc with low accretion dominated by the gas pressure (Ghosh \& Abramowicz 1997), which can be expressed as
\begin{equation}
\rm{L_{BZ,GPD}}=\rm{8\times10^{44}(\frac{M}{10^{8}M_{\odot}})^{11/10}(\frac{\dot{m}}{10^{-4}})^{4/5}j^{2}ergs^{-1}}
\end{equation}
where $\rm{\dot{m}}$ is the accretion rate.

\subsection{Jet power versus black hole mass and Eddington ratio}
Figure 3 shows jet power as a function of black hole mass (left panel) and jet power as a function of Eddington ratio (right panel). From Fig.3, we can see two different regimes. One is that jet power depends on the black hole mass and another is that jet power depends on the accretion. We find that the FSRQs and FRII-HEG are in the mass-dependent regime, while BL Lacs and FRI-LEG are in the accretion-dependent regime. In Fig.3, it is easy to recognize that BL Lacs and FRI-LEG are basically in the GPD regime, while FSRQs and FRII-HEG are in the RPD regime. Observational evidence that quasars can accelerate high-velocity winds is plentiful, Cattaneo et al.(2009) suggested that this winds is `momentum-drive' by radiation pressure. According to the SED modeling, Foschini (2011) also found that the FSRQs and $\gamma$-ray narrow line Seyfert1 are in RPD, while BL Lacs are basically in the GPD regime. We also find that the FSRQs and BL Lacs are placed along a line from low power/low accretion to high power/high accretion. This is well-known ``blazar main sequence'', where FSRQs have a strong disc and evolve to poorly accreting BL Lacs (Cavaliere \& D'Elia 2002; B$\ddot{o}$ttcher \& Dermer 2002; Maraschi \& Tavecchio 2003; Foschini 2011). We also note that some LBLs are in the mass-dependent region (left panel), which suggests that FSRQs and LBLs may have relationship. Li et al. (2010) have studied the relation between broadband spectral indices $\rm{\alpha_{ox}}$ and $\rm{\alpha_{x\gamma}}$ for Fermi blazars. They found that FSRQs and LBLs occupy the same region, which suggests that they have similar spectral properties.
\begin{figure*}[htbp]
\centerline{
\includegraphics[width=8.6cm,height=8cm]{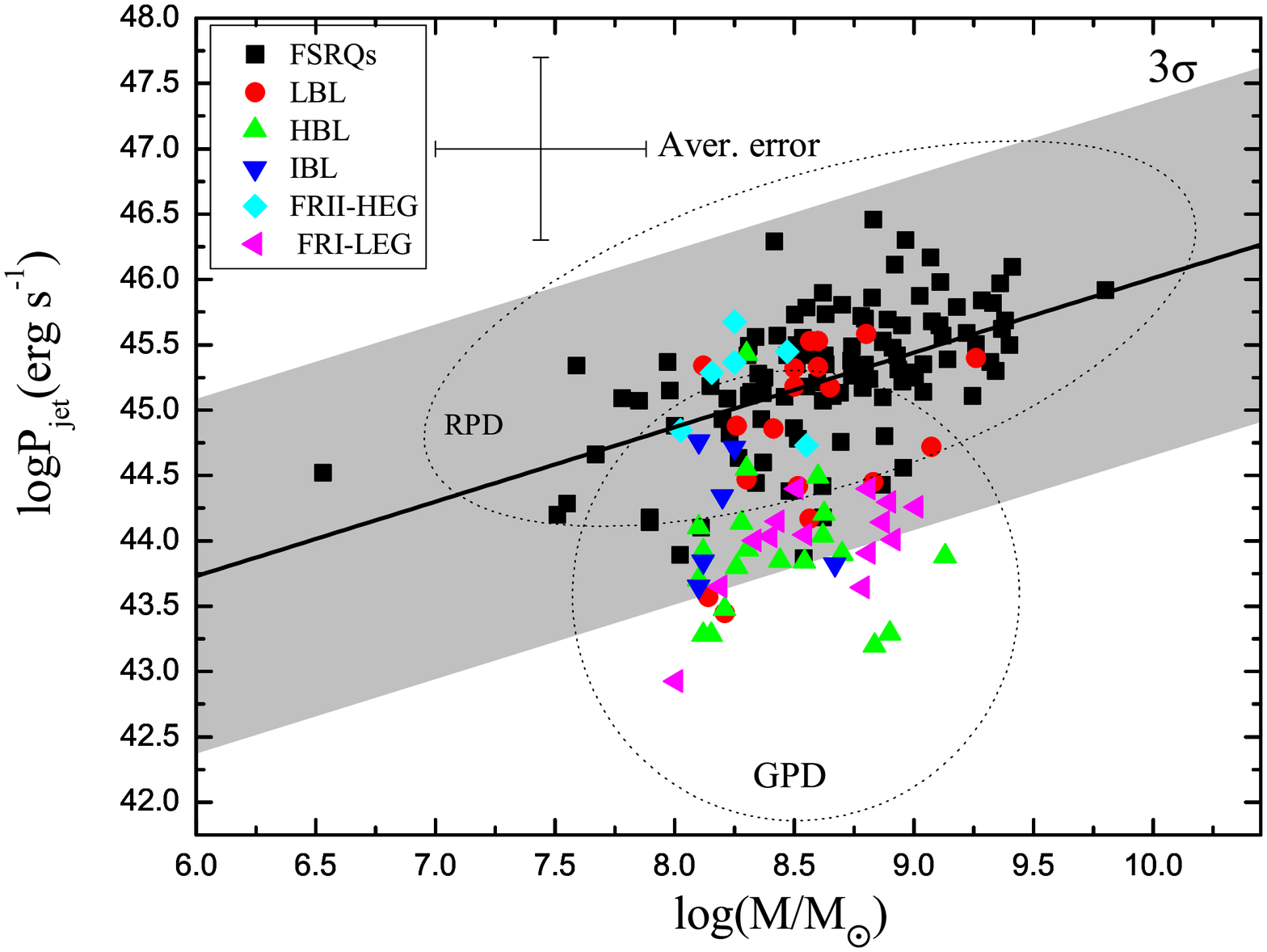}
\includegraphics[width=8.6cm,height=8cm]{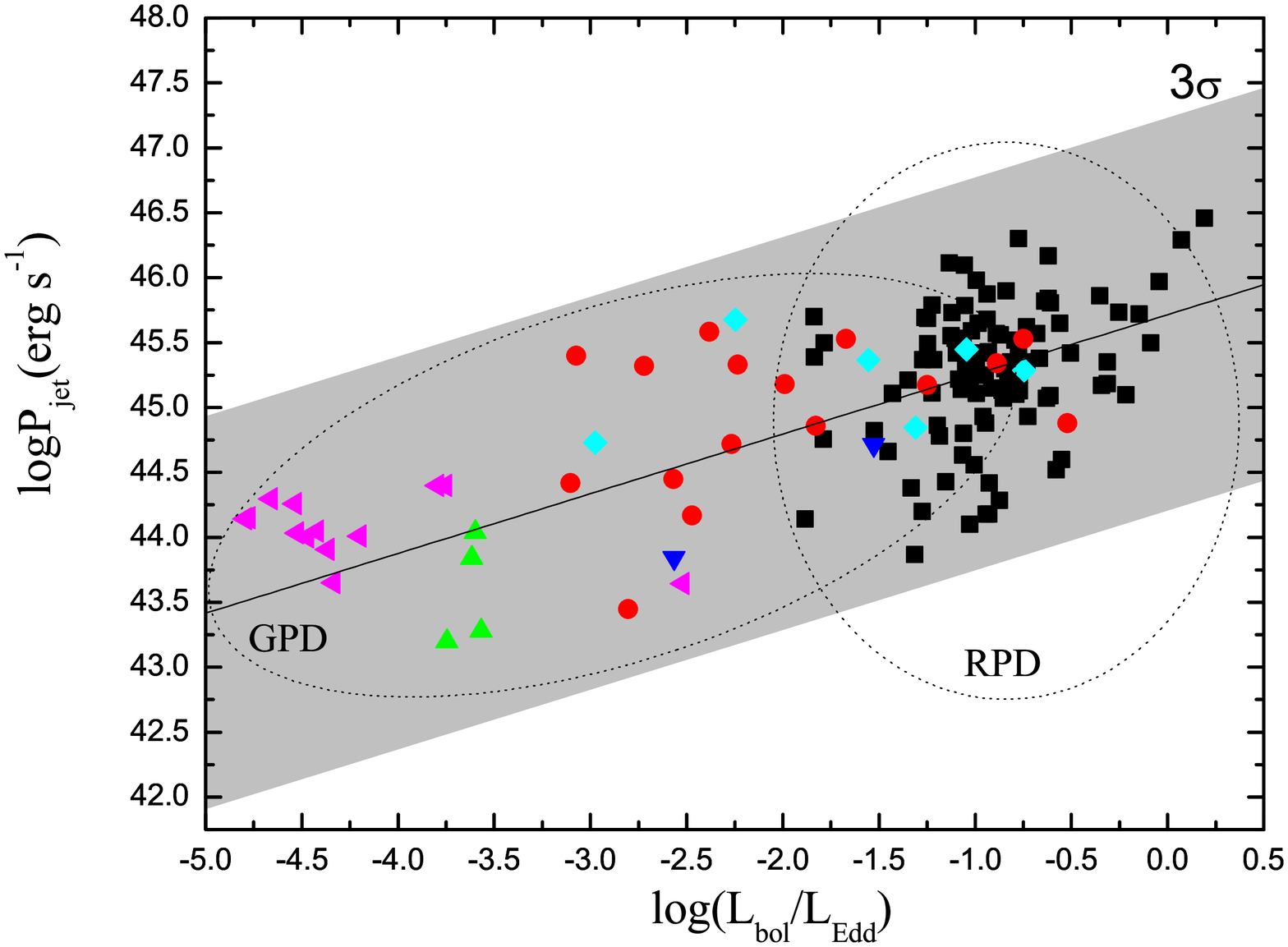}}
\caption{Jet power versus black hole mass (left panel) and jet power versus Eddington ratio (right panel), respectively. The 3$\sigma$ confidence band of the fits is shown as the grey shaded regions. The grey shaded regions indicate only the FSRQs and FRII-HEG in the $\rm{\log P_{jet}-\log M}$ plane; the grey shaded regions indicate all sources in the $\rm{\log P_{jet}-\log L_{bol}/L_{Edd}}$ plane. The black line is the best least square fits.}
\label{fig:opt}
\end{figure*}
\subsection{Ratio of jet power to BZ luminosity versus Eddington ratio}
We calculate the BZ luminosity by equation (3) and (4) for FSRQs/FRII-HEG and BL Lacs/FRI-LEG, respectively. Figure 4 shows the ratio of jet power to BZ luminosity is as a function of Eddington ratio. The maximum BZ luminosity of FSRQs/FRII-HEG and BL Lacs/FRI-LEG can be calculated by equation (3) and (4) with j=1, respectively. In Figure 4, we find that most of FSRQs/FRII-HEG and BL Lacs/FRI-LEG have $\rm{P_{jet}>L_{BZ}^{max}}$, which may indicate that the BZ mechanism is insufficient to explain the jet power of these objects. Foschini (2011) suggested that the Blandford-Znajek mechanism is sufficient to explain the power of BL Lacs. However, we find that our results are contrary to them. We notice that they only used 9 BL Lacs in their sample, while we use a large sample. At the same time, we also find that they used the upper limits on the broad line luminosity to the BL Lacs. These differences may lead to the different results. The above results may suggest that there is need to invoke alternative or additional mechanism to explain the jet power of these objects. A hybrid model (i.e., BP+BZ mechanism) has been recently adopted by Garofalo et al. (2010) to explain the observed differences in the AGN with relativistic jets. We also find a small number of FSRQs and BL Lacs have $\rm{P_{jet}<L_{BZ}^{max}}$, which may indicate that the BZ mechanism is sufficient to explain the jet power of this small number of FSRQs and BL Lacs.
\begin{figure}
\includegraphics[width=9cm,height=8cm]{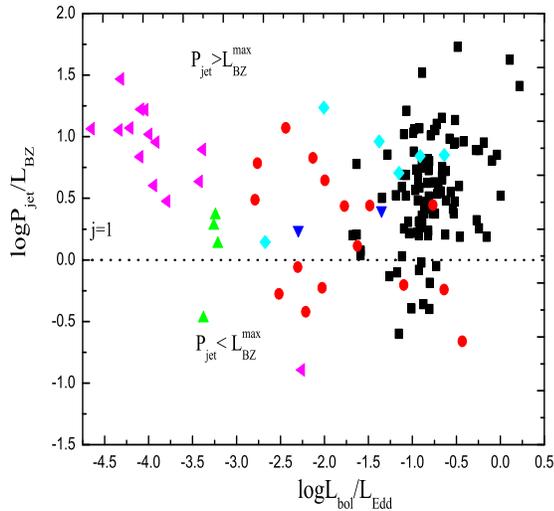}
\vspace{0pc}
\caption{Ratio between the jet power and the calculated luminosity according to the Blandford-Znajek theory. The dotted line is j=1. The meanings of different symbols are as same as Fig.3.}
\label{sample-fig2}
\end{figure}

\subsection{The FRI/FRII Dichotomy}
Meier (1999) given the dividing line between FRI and FRII sources of Ledlow and Owen (1996), which can be expressed as
\begin{equation}
\rm{\log{P_{rad}}}=\rm{-0.66M_{R}+10.35}
\end{equation}
where $\rm{P_{rad}}$ is the observed radio power at 1.4 GHz in units of W $\rm{Hz^{-1}}$ and $\rm{M_{R}}$ is the absolute optical R-band magnitude of the host galaxy. Mclure and Dunlop (2002) have gotten an empirical relation between $\rm{M_{R}}$ of the host galaxy and central BH mass, which can be expressed as
\begin{equation}
\rm{\log(M_{BH}/M_{\odot})}=\rm{-0.5(\pm0.2)M_{R}-2.96(\pm0.48)}
\end{equation}
Willott et al. (1999) pointed out that the jet power can be estimated by the following formula
\begin{equation}
\rm{Q_{jet}}\simeq\rm{3\times10^{38}f^{3/2}L_{151,ext}^{6/7}W}
\end{equation}
where $\rm{L_{151,ext}}$ is the extend radio luminosity at 151 MHz in units of $10^{28}$ W $\rm{Hz^{-1}sr^{-1}}$, and the normalization is uncertain and introduce the factor f. Blundell and Rawlings (2000) suggested that the factor f is most likely in the range of 10-20. Wu and Cao (2008) obtained the dividing line between FRI and FRII radio galaxies in the $\rm{Q_{jet}-M_{BH}}$ plane by using equation (5)-(7), Xu et al. (2009) have used different cosmological parameters that were presented in this Letter to modify it, which can be expressed as
\begin{equation}
\begin{split}
\rm{\log{Q_{jet}}(ergs^{-1})}=\rm{1.13\log{M_{BH}}(M_{\odot})}\\
{+33.18+1.5\rm{\log{f}}}
\end{split}
\end{equation}

In Figure 5, we plot the relation between the black hole mass (M) and jet power ($\rm{P_{jet}}$) for FSRQs and BL Lacs. Figure 5 shows that the FRI/FRII diving line was given by Ledlow and Owen (1996) roughly separates the FSRQs from BL Lacs in the $\rm{M-P_{jet}}$ plane with f=10/20, which supports unification schemes of AGN. Blundell and Rawlings (2000) suggested that $\rm{f\sim10}$ is likely the consequence of the evolution of magnetic field strength as radio sources evolve, which would mean that the bulk kinetic (jet) and radiative outputs of radio loud AGNs are similar magnitude (Willott et al. 1999). Willott et al. (1999) found that the ionization luminosity of radio galaxies is roughly equal to the jet power for f=20, which corresponds to $\rm{L_{ion}/L_{Edd}\sim2.5\times10^{-2}}$ for typical black hole mass $\rm{M_{BH}\sim10^{7.5-9.5}M_{\odot}}$ in sample of Ledlow and Owen (1996). It is still unclear why the jet power of FRII radio galaxies is always above this dividing line, which is beyond the scope of this work. From Fig.5, we also find that most of LBLs are above the dividing line, while most of HBLs are below the dividing line, which is similar to the FRI/FRII division. The HBLs have relatively lower jet power than LBLs and FSRQs. These results may provide the useful clues to investigate the relationship between FSRQs, LBLs and HBLs.
\begin{figure}
\includegraphics[width=9cm,height=8cm]{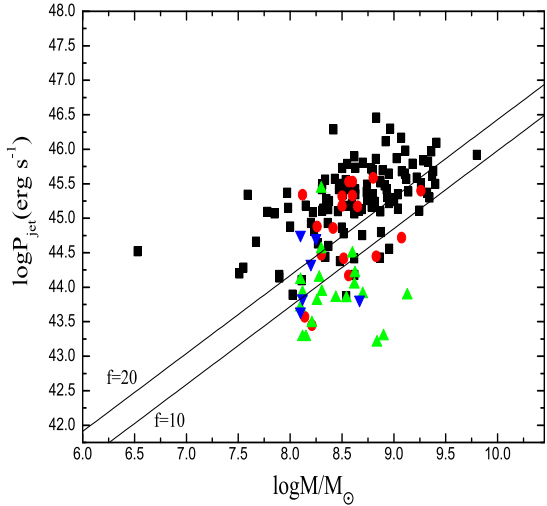}
\vspace{0pc}
\caption{Jet power as a function of black hole mass for various classes. The two solid line represent the Ledlow-Owen dividing line between FRI and FRII radio galaxies given by equation (8) with f=20[top] and f=10[bottom],respectively. The meanings of different symbols are as same as Fig.3.}
\label{sample-fig3}
\end{figure}
\begin{figure}
\includegraphics[width=9cm,height=8cm]{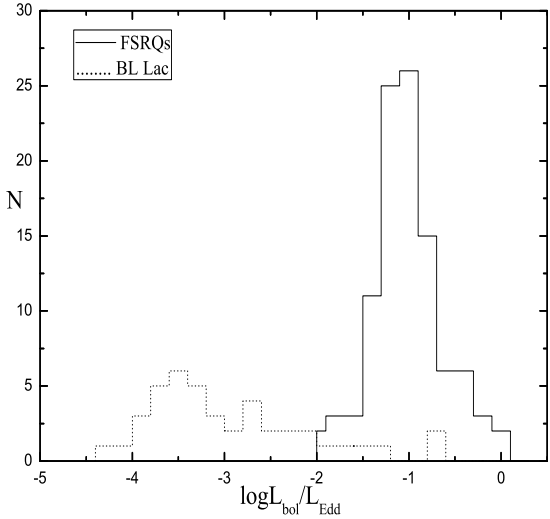}
\vspace{0pc}
\caption{Distributions of Eddington ratios for Fermi FSRQs (solid line) and BL Lac objects (dot line).}
\label{sample-fig4}
\end{figure}

In Figure 6, we show that the distributions of Eddington ratios for BL Lacs and FSRQs, which exhibit a bimodal nature. The FSRQs are clearly separated from the BL Lacs at $\rm{L_{bol}/L_{Edd}\sim0.01}$. Most of BL Lacs have $\rm{L_{bol}/L_{Edd}\leq0.01}$, while all FSRQs have $\rm{L_{bol}/L_{Edd}\geq0.01}$. These results that the bimodal behavior of the distribution may imply different accretion modes in BL Lacs and FSRQs. From Figure 6, we also find that the transition of accretion states happens at $\rm{L_{bol}/L_{Edd}\sim0.01}$. Narayan and Yi (1995) suggested that the ADAF may be presented when $\rm{L_{bol}/L_{Edd}\leq0.01}$, which suggested that ADAFs are presented in BL Lacs and standard thin disks are in FSRQs. A similar explanation is proposed to explain the FRI/FRII division, in which ADAFs would be presented in FRI radio galaxies and the standard thin disks are in FRII radio galaxies (e.g., Ghisellini \& Celotti 2001; Wu \& Cao 2008; Xu et al. 2009).
\subsection{The jet power vs broad line luminosity}
The properties of relativistic jets are thought to be closely linked with the properties of both the accretion disk and the black hole in AGNs. However, the origin and formation of relativistic jets are still an unsolved mystery in astrophysics. In current theoretical models of the formation of jet, power is generated via accretion and extraction of rotational energy of disc/black hole (Blandford \& Znajek 1977; Blandford \& Payne 1982), and then converted into the kinetic power of the jet. The relation between jet power and broad line luminosity is shown in Figure 7. We use a linear regression to analyze the relationship between jet power and broad line luminosity. From figure 7, we find a significant correlation between them (r=0.7081, P$<$0.0001). This result supports the close connection between jet power and accretion. The result is consistent with other authors (e.g., Rawlings \& Saunders 1991; Falcke \& Biermann 1995; Cao \& Jiang 1999; Sbarrato et al. 2012,2014; Xiong \& Zhang 2014). According to SED modeling, Ghisellini et al. (2014) also found a close connection between jet power and accretion for Fermi blazars.
\begin{figure}
\includegraphics[width=9cm,height=8cm]{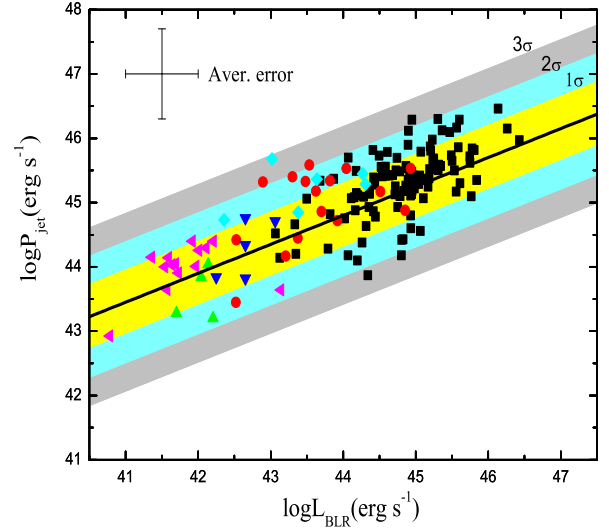}
\vspace{0pc}
\caption{Jet power as a function of broad line luminosity for various classes. Shaded colored areas correspond to 1, 2 and 3 $\sigma$ (vertical) dispersion, $\sigma=0.5$ dex. The black line is best least square fits ($\rm{\log P_{jet}=0.45\log L_{BLR}+25}$). The meanings of different symbols are as same as Fig.3.}
\label{sample-fig4}
\end{figure}
\section{CONCLUSIONS}
In this work, we have explored the relationship between FSRQs, BL Lacs, FRI-LEG and FRII-HEG radio galaxies. Our main results are the following:

(i) According to study the $\rm{Q_{jet}-M}$ and $\rm{Q_{jet}-L_{bol}/L_{Edd}}$ plane, we find two different regimes. One is the jet power depends on the accretion and another is the jet power depends on the black hole mass. FSRQs and FRII-HEG are in the RPD regime (mass-dependent regime), while BL Lacs and FRI-LEG are in the GPD regime (accretion-dependent regime).

(ii) Through studying the ratio of between the jet power and the calculated the luminosity according to the Blandford-Znajek theory versus the Eddington ratios plane, we find that the most of both FSRQs and FRII-HEG and both BL Lacs and FRI-LEG have $\rm{P_{jet}>L_{BZ}^{max}}$, which suggests that the BZ mechanism is insufficient to explain the jet power of these objects. There is need to invoke alternative or additional mechanism to explain the jet power of these objects. We also find a small number of FSRQs and BL Lacs have $\rm{P_{jet}<L_{BZ}^{max}}$, which may indicate that the BZ mechanism is sufficient to explain the jet power of this small number of FSRQs and BL Lacs.

(iii) We find that the FSRQs are roughly separated from the BL Lacs by the Ledlow-Owen FRI/FRII dividing line in the $\rm{M-P_{jet}}$ plane. This result supports the unification schemes of AGNs.

(iv) The Eddington ratios $\rm{L_{bol}/L_{Edd}}$ of BL Lacs are systematically lower than those of FSRQs in our sample with a clearly division at $\rm{L_{bol}/L_{Edd}\sim0.01}$. The Eddington distributions of BL Lacs and FSRQs exhibit a bimodal nature, which imply that the accretion mode of BL Lacs may be different from that of FSRQs.

(v) We find a significant correlation between jet power and broad line luminosity in AGNs, which supports the close connection between jet power and accretion.
\acknowledgements
We sincerely thank the anonymous referee for valuable
comments and suggestions. We are very grateful to the Science
Foundation of Yunnan Province of China(2012FB140,2010CD046). This work is supported by
the National Nature Science Foundation of China (11063004,11163007,U1231203), and the High-Energy Astrophysics Science and Technology Innovation Team of Yunnan Higher School and Yunnan Gravitation Theory Innovation Team (2011c1). This research has made use of the NASA/IPAC Extragalactic Database (NED), that is operated by Jet Propulsion Laboratory, California Institute of Technology, under contract with the National Aeronautics and
Space Administration.


\begin{thebibliography}{99}
\bibitem[Abdo et al. (2010a)]{Abd0a} Abdo A. A., Ackermann M., Agudo I., et al., 2010a, ApJ, 716, 30
\bibitem[Abdo et al. (2010)]{Abd10} Abdo A.A., et al. 2010, ApJS, 188, 405
\bibitem[Abdo et al. (2009)]{Abd09} Abdo A.A., et al., 2009, arXiv: 0912.0973
\bibitem[Abdo et al. (2009a)]{Abd9a} Abdo A.A., et al., 2009a, ApJ, 707, L142
\bibitem[Ackermann et al. (2011)]{Ack11} Ackermann M., Ajello M., Allafort A., et al., 2011, ApJ, 743, 171
\bibitem[Bai et al. (2001)]{Bai01} Bai J. M., \& Lee M. G. 2001, ApJ, 548, 244
\bibitem[Baum et al. (1995)]{Bau95} Baum S. A., Zirbel E. L., \& O'Dea C. P., 1995, ApJ, 451, 88
\bibitem[Best et al. (2012)]{Bes12} Best, P.N., \& Heckman, T.M., 2012, MNRAS, 421, 1569
\bibitem[Bicknell et al. (1995)]{Bic95} Bicknell G.V., 1995, ApJS, 101, 29
\bibitem[Blandford et al. (1999)]{Bla99} Blandford R. D., \& Begelman M. C., 1999, MNRAS, 303, L1
\bibitem[Blandford et al. (1982)]{Bla82} Blandford R.D., \& Payne D.G., 1982, MNRAS, 199,883
\bibitem[Blandford et al. (1977)]{Bla77} Blandford R.D., \& Znajek R.L., 1977, MNRAS, 179, 433
\bibitem[Blandford et al. (1978)]{Bla78} Blandford R.D., Rees M.J., 1978, in Pittsburgh Conference on BL Lac Objects. Pittsburgh Univ., PA, p.328
\bibitem[Blundell et al. (2000)]{Blu00} Blundell K. M., \& Rawlings, S. 2000, AJ, 119, 1111
\bibitem[Buttiglione et al. (2009)] {But09} Buttiglione S., Capetti A., Celotti A., Axon D. J., Chiaberge M.,Macchetto F. D., SparksW. B., 2009, A\&A, 495, 1033
\bibitem[Buttiglione et al. (2010)] {But10} Buttiglione S., Capetti A., Celotti A., Axon D. J., Chiaberge M.,Macchetto F. D., SparksW. B., 2010, A\&A, 509, 6
\bibitem[B$\ddot{o}$ttcher et al. (2002)]{Bot02} B$\ddot{o}$ttcher M., \& Dermer C.D., 2002, ApJ, 564, 86
\bibitem[Cao et al. (1999)]{Cao99} Cao X.,\& Jiang D.R., 1999, MNRAS, 307, 802
\bibitem[Cao et al. (2002)]{Cao02} Cao X., 2002, ApJ, 570, L13
\bibitem[Cao et al. (2003)]{Cao03} Cao X., 2003, ApJ, 599, 147
\bibitem[Cattaneo et al.(2009)]{Cat09} Cattaneo,A., et al., 2009,Nature,460,213
\bibitem[Capetti et al. (2000)]{Cap00} Capetti A., Trussoni E., Celotti A., Feretti L., \& Chiaberge M. 2000, MNRAS,318,493
\bibitem[Cavagnolo et al. (2010)]{Cav10} Cavagnolo K.W. et al., 2010, ApJ, 720, 1066
\bibitem[Cavaliere et al. (2002)]{Cav02} Cavaliere A., \& D'Elia V., 2002, ApJ, 571, 226
\bibitem[Celotti et al. (1997)]{Cel97} Celotti A., Padovani P., \& Ghisellini G., 1997, MNRAS, 286, 415
\bibitem[Chai et al. (2012)]{Cha12} Chai B., Cao X., \& Gu M., 2012, ApJ, 759, 114
\bibitem[Dunlop et al. (2003)]{Dun03} Dunlop J.S., McLure R.J., Kukula M.J., et al., 2003, MNRAS, 340, 1095
\bibitem[Fanaroff et al. (1974)]{Fan74} Fanaroff B.L, \& Riley J.M., 1974, MNRAS, 167, 31
\bibitem[Falcke et al. (1995)]{Fal95} Falcke, H., \& Biermann, P.L., 1995, A\&A, 293, 665
\bibitem[Ferrarese et al. (2001)]{Fer01} Ferrarese,L., et al., 2001, ApJ,555, L79
\bibitem[Foschini et al. (2011)]{Fos11} Foschini F., 2011, RAA, 11,1266
\bibitem[Francis et al. (1991)]{Fra91} Francis P.J., Hewett P.C., Foltz C.B., Chaffee F.H., Wey-mann R.J. \& Morris S.L., 1991, ApJ, 373, 465
\bibitem[Garofalo et al. (2010)]{Gar10} Garofalo D., et al., 2010, MNRAS, 406, 975
\bibitem[Gebhardt et al. (2000)]{Geb00} Gebhardt,K., et al., 2000, ApJ, 543, L5
\bibitem[Gendre et al. (2013)]{Gen13} Gendre, M.A., Best, P.N., Wall,J.V.,\& Ker,L.M.,2013, MNRAS,430,3086
\bibitem[Ghisellini et al. (2001)]{Ghi01} Ghisellini, G., \& Celotti, A., 2001, A\&A, 379, L1
\bibitem[Ghisellini et al. (2009)]{Ghi09} Ghisellini, G., Maraschi L. \& Tavecchio F., 2009, MNRAS, 396, L105
\bibitem[Ghisellini et al. (2010)]{Ghi10} Ghisellini, G., Tavecchio, F., Foschini, L., Ghirlanda, G., Maraschi, L., \& Celotti, A., 2010, MNRAS, 402, 497
\bibitem[Ghisellini et al. (2014)]{Ghi14} Ghisellini1,G., Tavecchio1,F., Maraschi,L., Celotti,A.,\& Sbarrato, T., 2014, Nature, 515,376
\bibitem[Ghosh et al. (1997)]{Gho97} Ghosh P., \& Abramowicz M.A., 1997, MNRAS, 292, 887
\bibitem[Gopal-Krishna et al. (2000)] {Gop03} Gopal-Krishna Wiita, P. J., 2000, A\&A, 363, 507
\bibitem[Gu et al. (2001)]{Gu001} Gu M., Cao, X.W.., \& Jiang, D.R., 2001, MNRAS, 327, 1111
\bibitem[Hardcastle et al. (2007)]{Har07} Hardcastle M. J., Evans D. A., \& Croston J. H., 2007, MNRAS, 376, 1849
\bibitem[Heckman et al. (2014)]{Hec14} Heckman T.M., \& Best P.N., 2014, ARA\&A, 52,589
\bibitem[Laing et al. (1994)]{Lai94} Laing R. A., Jenkins C. R., Wall J. V., Unger S. W., 1994, in Bicknell G. V., Dopita M. A., Quinn P. J., eds, ASP Conf. Ser. Vol. 54, The First Stromlo Symp. The Physics of Active Galaxies. Astron. Soc. Pac., San Francisco, p. 201
\bibitem[Landt et al. (2004)]{Lan04} Landt H., Padovani P., Giommi P., \& Perlman E.S., 2004, MNRAS, 351, 83
\bibitem[Ledlow et al. (1996)]{Led96} Ledlow M. J., \& Owen F. N., 1996, AJ, 112, 9
\bibitem[Leon-Tavares et al. (2011a)]{Leo1a} Leon-Tavares J. et al., 2011a, MNRAS, 411, 1127
\bibitem[Leon-Tavares et al. (2011b)]{Leo1b} Leon-Tavares J. et al., 2011b, A\&A, 532, 146
\bibitem[Li et al. (2010)]{Li010} Li H.Z., Xie G.Z., \& Chen L.E., 2010, ApJ, 709, 1407
\bibitem[Liu et al. (2006)]{Liu06} Liu Y., Jiang D.R., \& Gu M.F., 2006, ApJ, 637, 669
\bibitem[Maraschi et al. (2003)]{Mar03} Maraschi L., \& Tavecchio F., 2003, ApJ, 593, 667
\bibitem[Marchesini et al. (2004)]{Mar04} Marchesini, D., Celotti A., \& Ferrarese L., 2004, MNRAS, 351, 733
\bibitem[McLure et al. (2001)]{McL01} McLure,R.J., \& Dunlop,J.S., 2001, MNRAS, 327, 199
\bibitem[McLure et al. (2011)]{McL11} McLure R., \& Dunlop 2002, MNRAS, 331, 795
\bibitem[Meier et al. (1999)]{Mei99} Meier D. L. 1999, ApJ, 522, 753
\bibitem[Meier et al. (2001)]{Mei01} Meier D., Koide S., \& Uchida Y., 2001, Science, 291, 84
\bibitem[Meyer et al. (2011)]{Mey11} Meyer E.T., et al., 2011, ApJ, 740,98
\bibitem[Narayan et al. (1997)]{Nar97} Narayan R., Garcia M. R., \& McClintock J. E., 1997, ApJ, 478, L79
\bibitem[Nemmen et al. (2012)]{Nem12} Nemmen R.S., Georganopoulos M., Guiriec S., Meyer E.T., Gehrels N., \& Sambruna R.M., 2012, Science, 338, 1445
\bibitem[Owen et al. (1996)]{Owe96} Owen F. N., Ledlow M. J., \& Keel W. C. 1996, AJ, 111, 53
\bibitem[Padovani et al. (1991)]{Pad91} Padovani P., \& Urry C.M. 1991, ApJ, 368, 373
\bibitem[Padovani et al. (1992)]{Pad92} Padovani P., \& Urry C.M. 1992, ApJ, 387, 449
\bibitem[Rawlings et al. (1991)]{Raw91} Rawlings,S., \& Saunders, R., 1991, Nature, 349, 138
\bibitem[Rees et al. (1982)]{Ree82} Rees M. J., Begelman M. C., Blandford R. D., \& Phinney E. S., 1982, Nat,295, 17
\bibitem[Reynolds et al. (1996)]{Rey96} Reynolds C. S., di Matteo T., Fabian A. C., Hwang U., \& Canizares C. R.,1996, MNRAS, 283, L111
\bibitem[Sbarrato et al. (2012)]{Sba12} Sbarrato T., Ghisellini G., Maraschi L., \& Colpi M., 2012, MNRAS, 421, 1764
\bibitem[Sbarrato et al. (2014)]{Sba14} Sbarrato T., Padovani P., \& Ghisellini G., 2014, MNRAS, 445, 81
\bibitem[Shakura et al. (1973)]{Sha73} Shakura N. I., \& Sunyaev R. A., 1973, A\&A, 24, 337
\bibitem[Shaw et al. (2012)]{Sha12} Shaw M.S., Romani R.W., Cotter G. et al., 2012, ApJ, 748, 49
\bibitem[Shen et al. (2011)]{She11} Shen Y., Richards G.T., Strauss M.A. et al., 2011, ApJS, 194, 45
\bibitem[Tremaine et al. (2002)]{Tre02} Tremaine,S., et al., 2002, ApJ, 574, 740
\bibitem[Urry et al. (1995)]{Urr95} Urry C.M., \& Padovani P., 1995, PASP, 107, 803
\bibitem[Wang et al. (2004)]{Wan04} Wang J.M., Luo B., \& Ho L.C., 2004, ApJ, 615, L9
\bibitem[Willott et al. (1999)]{Wil99} Willott, C. J., Rawlings, S., Blundell, K. M., \& Lacy, M., 1999, MNRAS, 309, 1017
\bibitem[Woo et al. (2002)]{Woo02} Woo J.H., \& Urry C.M., 2002, ApJ, 579, 530
\bibitem[Wu et al. (2011)]{Wu011} Wu Q.W., Cao X.W., \& Wang D.X., ApJ, 2011,735, 50
\bibitem[Xie et al. (2007)]{Xie07} Xie G.Z., Dai H., \& Zhou S.B., 2007, AJ, 134, 1464
\bibitem[Xie et al. (1991)]{Xie91} Xie G.Z., et al., 1991, A\&A, 249, 65
\bibitem[Xie et al. (2004)]{Xie04} Xie G.Z., Zhou S.B., \& Liang E.W., 2004, AJ, 127, 53
\bibitem[Xiong et al. (2014)]{Xio14} Xiong D.R., \& Zhang X., 2014, MNRAS, 441, 3375
\bibitem[Xu et al. (2009)]{Xu009} Xu Y.D., Cao X.W., \& Wu Q.W., 2009, ApJ, 694, L107
\bibitem[Zhang et al. (2012)]{Zha12} Zhang J., Liang E.W., Zhang S.N., \& Bai J.M., 2012, ApJ, 752, 157
\bibitem[Zhou et al. (2009)]{Zho09} Zhou M., \& Cao X., 2009, RAA, 9, 293
\bibitem[Zirbel et al. (1995)]{Zir95} Zirbel, E.L., \& Baum, S.A., 1995, ApJ, 448, 521
\end{thebibliography}
\end{document}